## Journal of Geophysical Research: Space Physics

# Energetic electron precipitation into the middle atmosphere—Constructing the loss cone fluxes from MEPED POES

H. Nesse Tyssøy[1], M. I. Sandanger[1], L.-K. G. Ødegaard[1], J. Stadsnes[1], A. Aasnes[1], and A. E. Zawedde[1]

[1]Birkeland Centre for Space Science, Department of Physics and Technology, University of Bergen, Bergen, Norway

**Abstract** The impact of energetic electron precipitation (EEP) on the chemistry of the middle atmosphere (50–90 km) is still an outstanding question as accurate quantification of EEP is lacking due to instrumental challenges and insufficient pitch angle coverage of current particle detectors. The Medium Energy Proton and Electron Detectors (MEPED) instrument on board the NOAA/Polar Orbiting Environmental Satellites (POES) and MetOp spacecraft has two sets of electron and proton telescopes pointing close to zenith (0°) and in the horizontal plane (90°). Using measurements from either the 0° or 90° telescope will underestimate or overestimate the bounce loss cone flux, respectively, as the energetic electron fluxes are often strongly anisotropic with decreasing fluxes toward the center of the loss cone. By combining the measurements from both telescopes with electron pitch angle distributions from theory of wave-particle interactions in the magnetosphere, a complete bounce loss cone flux is constructed for each of the electron energy channels $>50$ keV, $>100$ keV, and $>300$ keV. We apply a correction method to remove proton contamination in the electron counts. We also account for the relativistic ($>1000$ keV) electrons contaminating the proton detector at subauroral latitudes. This gives us full range coverage of electron energies that will be deposited in the middle atmosphere. Finally, we demonstrate the method's applicability on strongly anisotropic pitch angle distributions during a strong geomagnetic storm in February 2008. We compare the electron fluxes and subsequent energy deposition estimates to OH observations from the Microwave Limb Sounder on the Aura satellite substantiating that the estimated fluxes are representative for the true precipitating fluxes impacting the atmosphere.

## 1. Introduction

The almost continuous precipitation of low-energy auroral electrons ($<30$ keV) from the magnetospheric plasmasheet deposit their energy above 90 km in the auroral oval. The auroral electrons are often isotropic, and a positive correlation between geomagnetic activity or particle precipitation and the $NO_x$ composition is fairly well established [*Baker et al.*, 2001; *Barth et al.*, 2003; *Sætre et al.*, 2004, 2006; *Hendrickx et al.*, 2015]. The role of the more energetic electron precipitation (EEP) in changing the chemistry of the middle atmosphere (50–100 km) is, on the other hand, still an outstanding question [*Rozanov et al.*, 2012; *Sinnhuber et al.*, 2012]. These often highly anisotropic electrons originate from the outer radiation belts, and their precipitation to the atmosphere is related to complex wave-particle interactions mostly confined to subauroral latitudes. The relation of EEP events to geomagnetic activity is not well defined yet, but it is known to be associated with both enhanced solar wind pressure caused by coronal mass ejections and high-speed solar wind streams (HSSWS) [*Turunen et al.*, 2009; *Meredith et al.*, 2011]. As the different events might have quite different geomagnetic signatures, recent studies apply energetic electron measurements to indicate the coexistence of electron precipitation and atmospheric chemical changes. For example, *Newnham et al.* [2011], *Daae et al.* [2012], and *Andersson et al.* [2012, 2014a, 2014b] suggest that electron precipitation during moderate but frequently occurring geomagnetic storms may have a significant impact on the $HO_x$ and $NO_x$ production rate and subsequently the catalytic destruction of ozone.

Accurate quantification of the effect of energetic electron precipitation, however, remains due to instrumental challenges. Most of the current particle detectors in space are unsuitable for determining the amount of particles precipitating into the atmosphere. The majority measures only the trapped particle fluxes because of inadequate pitch angle resolution [*Rodger et al.*, 2013]. In this respect, the Medium Energy Proton and Electron Detectors (MEPED) on board the Polar Orbiting Environmental Satellites (POES) and European Organisation for the Exploitation of Meteorological Satellites (EUMETSAT) MetOp has an expedient design.

MEPED consist of two electron and two proton telescopes, pointed in two directions, approximately 0° and 90° to the local vertical. At middle and high latitudes the 0° telescope measures particle fluxes that will be lost to the atmosphere, whereas the 90° telescope might detect precipitating particle fluxes and/or trapped particles in the radiation belts [Rodger et al., 2010]. The long measurement time series and multiple local time coverage make the POES/MEPED and MetOp/MEPED data set valuable in respect to estimating the medium- to high-energy particle loss to the atmosphere.

The pitch angle, $\alpha$, of a charged particle trapped in the magnetosphere is defined as the angle between its velocity vector and the magnetic field line it gyrates around. While the pitch angle changes as the particle moves along the magnetic field, the equatorial pitch angle, $\alpha_{eq}$, is often used as a reference. Here the particles may have a range of different pitch angle distributions (PADs). The level of anisotropy varies significantly with particle energy, location, and geomagnetic activity. This implies that the 0° and 90° telescopes cannot alone be used to determine the level of precipitating particle fluxes. Only in a rare case of strong pitch angle diffusion and an isotropic distribution will the 0° or 90° telescope give a realistic precipitating flux estimate. In case of an anisotropic distribution, the 0° detector will underestimate, while the 90° detector will overestimate the flux of precipitating electrons.

To overcome this challenge, Rodger et al. [2013] applied a geometric mean between the fluxes measured by the 0° and 90° detectors to determine the true precipitating electron fluxes. They examined the EEP measurement during ~250 satellite overflights of the Kilpisjärvi riometer in Finland ($L = 6.13$). As the riometer responds to the precipitating fluxes, an estimate of the energy deposition and its subsequent ionization and the expected riometer absorption can be used to compare the different measurements. At high flux levels ($> 10^6 \, cm^{-2} \, s^{-1} \, sr^{-1}$ for $>30 \, keV$) there was found a relatively good agreement between the absorption estimated from the geometric mean of the flux measurement and the riometer absorption. At lower flux levels, however, the space-based measurements significantly underestimate the absorption by ~7–9 times. In fact, the level of agreement between a geometric mean of the space based electron fluxes and the riometer absorption appears to be strongly dependent upon the level of diffusion and hence the level of isotropy in the PADs. Considering that the Kilpisjärvi riometer is located in the auroral zone where isotropy is more common compared to the subauroral latitudes where the medium energy (100–300 keV) and relativistic (>300 keV) electron precipitation usually occur, the geometric mean will not provide sufficient accuracy in determining the particle energy deposition.

The level of diffusion itself has previously been determined from the MEPED POES observations [Kirkwood and Osepian, 2001; Li et al., 2013, 2014a, 2014b] with the purpose of identifying the sources for pitch angle scattering such as chorus, hiss, and whistler mode waves. The 0° and 90° fluxes were fitted onto the solution of the Fokker-Planck equation for wave-particle diffusion [Kennel and Petschek, 1966]. Although, the Fokker-Planck equation requires steady state conditions where the particles being lost will be replenished so that the pitch angle distribution is in diffusion equilibrium, Li et al. [2013] and multiple follow-up papers have shown that without any strict criteria the diffusion coefficient estimated from the MEPED instrument was consistent with independent observations of chorus waves by the Van Allen Probes.

Estimating the diffusion strength from the POES/MEPED 0° and 90° detector forms the base of our study. After accounting for proton degradation [Sandanger et al., 2015; Ødegaard et al., 2016] and electron contamination [Evans and Greer, 2000; Yando et al., 2011], we identify the level of diffusion and hence the angular distribution within the loss cone. We take into account the detector sensitivity when the directional flux varies over the field of view solid angle of the telescope. In addition, we estimate the flux of relativistic electrons (>1000 keV) which contaminate the highest MEPED proton channel [Yando et al., 2011]. In total, these steps make up a complete tool box for overcoming the known challenges in regard to proton contamination, degradation, and insufficient loss cone information for the MEPED electron data, giving us a unique estimate of the precipitating electron fluxes in the mesosphere. Finally, we use the observed fluxes during a HSSWS demonstrating its applicability. Comparing the estimated fluxes and subsequent energy deposition to the OH density as measured by the Microwave Limb Sounder (MLS) on board the Aura satellite, we substantiate that our method gives a realistic estimate of precipitating energetic electron fluxes. To our knowledge, this is the first time direct satellite electron measurements are used to quantitatively determine its impact on $HO_x$ during a weak geomagnetic storm.

## 2. The MEPED Instrument and Data

MEPED is part of the Space Environment Monitor 2 (SEM-2) instrument package on board the POES and MetOp satellites, which are polar orbiting Sun-synchronous satellites at an altitude of ~850 km with an orbital period of ~100 min. MEPED includes an omnidirectional detector system that detects protons of higher energies than 16 MeV, but our focus is the four solid-state directional telescopes that measures electron fluxes in three energy ranges >50 keV, >100 keV, and >300 keV, and proton fluxes in five differential channels from 30 to 6900 keV and one integral channel >6900 keV.

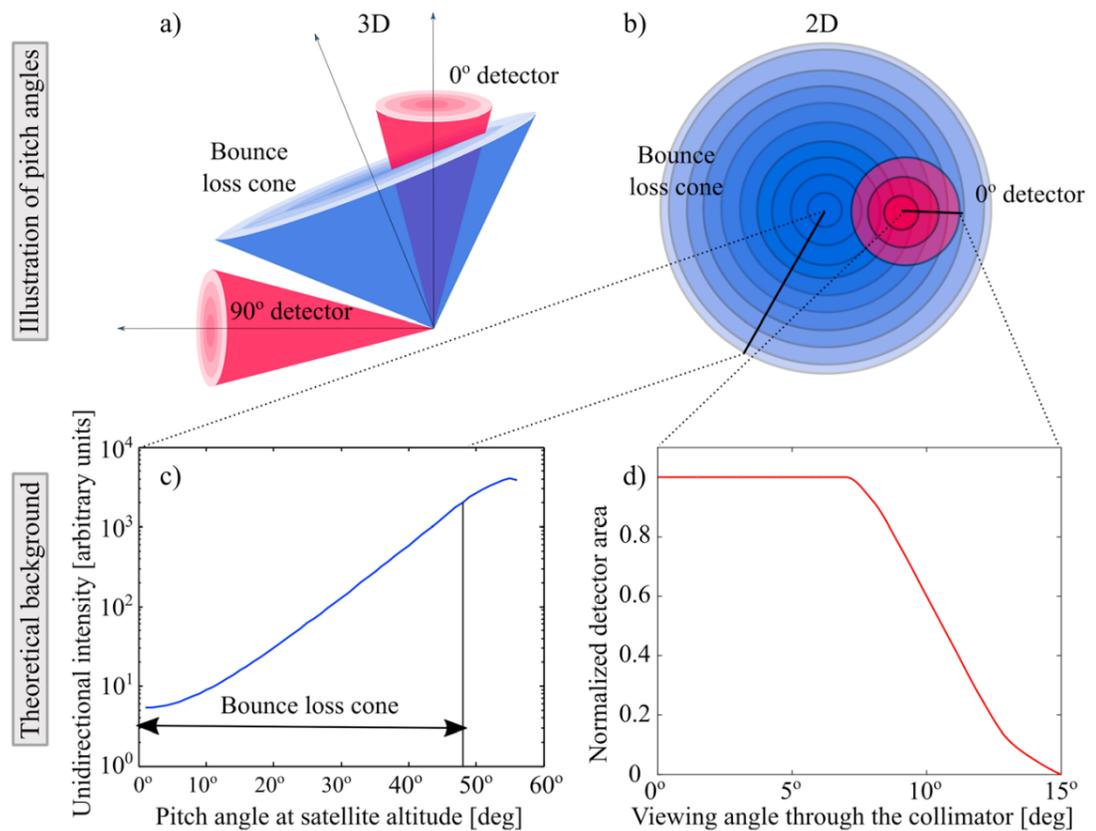

**Figure 1.** (a and b) The MEPED telescope viewing angles in respect to the loss cone exemplified for one case at high latitudes. (c) A theoretical pitch angle distribution profile based on a selected solution of the Fokker-Planck equation for particles. (d) The detector response-function for different viewing angles through the detector collimator.

The field of view of both the telescopes is 30° full width, with a higher detector response at the center angles as illustrated in Figure 1. The MEPED raw data are sampled every other second with 1 s integration period [*Evans and Greer*, 2000]. We average the fluxes over a 16 s interval, which corresponds to about 100 km along the satellite track or approximately 1° latitude similar to *Codrescu et al.* [1997]. The MEPED data files also list the pitch angle estimated from the International Geomagnetic Reference Field (IGRF) model, a parameter necessary to determine where the telescopes are pointing relative to the loss cone.

## 3. Proton Correction of the Electron Fluxes

The MEPED electron data are not straight forward to use. The electron telescopes are known to exhibit a spurious response to protons [*Evans and Greer*, 2000; *Yando et al.*, 2011] as listed in Table 1. In order to correct for the false counts, we can use the proton flux measurement from the proton telescopes.

It has, however, been well documented that the solid state detectors will degrade over time as a result of radiation damage [*Galand and Evans*, 2000; *Asikainen and Mursula*, 2011; *Asikainen et al.*, 2012; *Sandanger et al.*, 2015]. This impact is significant after 2–3 years of operation, changing the energy ranges of the proton detector. The degradation needs to be taken into account in a quantitative assessment of the data. The first effort to derive correction factors for the new energy ranges for the proton telescopes on board the different satellites was carried out by *Asikainen and Mursula* [2011] and followed up by *Asikainen et al.* [2012] and *Ødegaard* [2013]. Recently, a new study by *Sandanger et al.* [2015] uses a larger statistical database showing consistent result, indicating a robust calibration procedure. A follow-up study by *Ødegaard et al.* [2016] shows that all detectors degrade in the same manner in the same phase of a solar cycle. This knowledge is applied to determine correction factors to create a complete set of correction factors for the SEM-2 package up to present time.

**Table 1.** Summary of Electron Energy Channels and Their Sensitivity to Electron and Proton Fluxes

| Energy Channel | Electron Energy Range | Contaminating Protons |
| --- | --- | --- |
| E1 | >50 keV[a] | 210–2600 keV |
| E2 | >100 keV | 280–2600 keV |
| E3 | >300 keV | 440–2600 keV |
| P6 | >1000 keV[a] | |

[a]Effective energy range derived from Geant-4 geometric factors [*Yando et al.*, 2011].

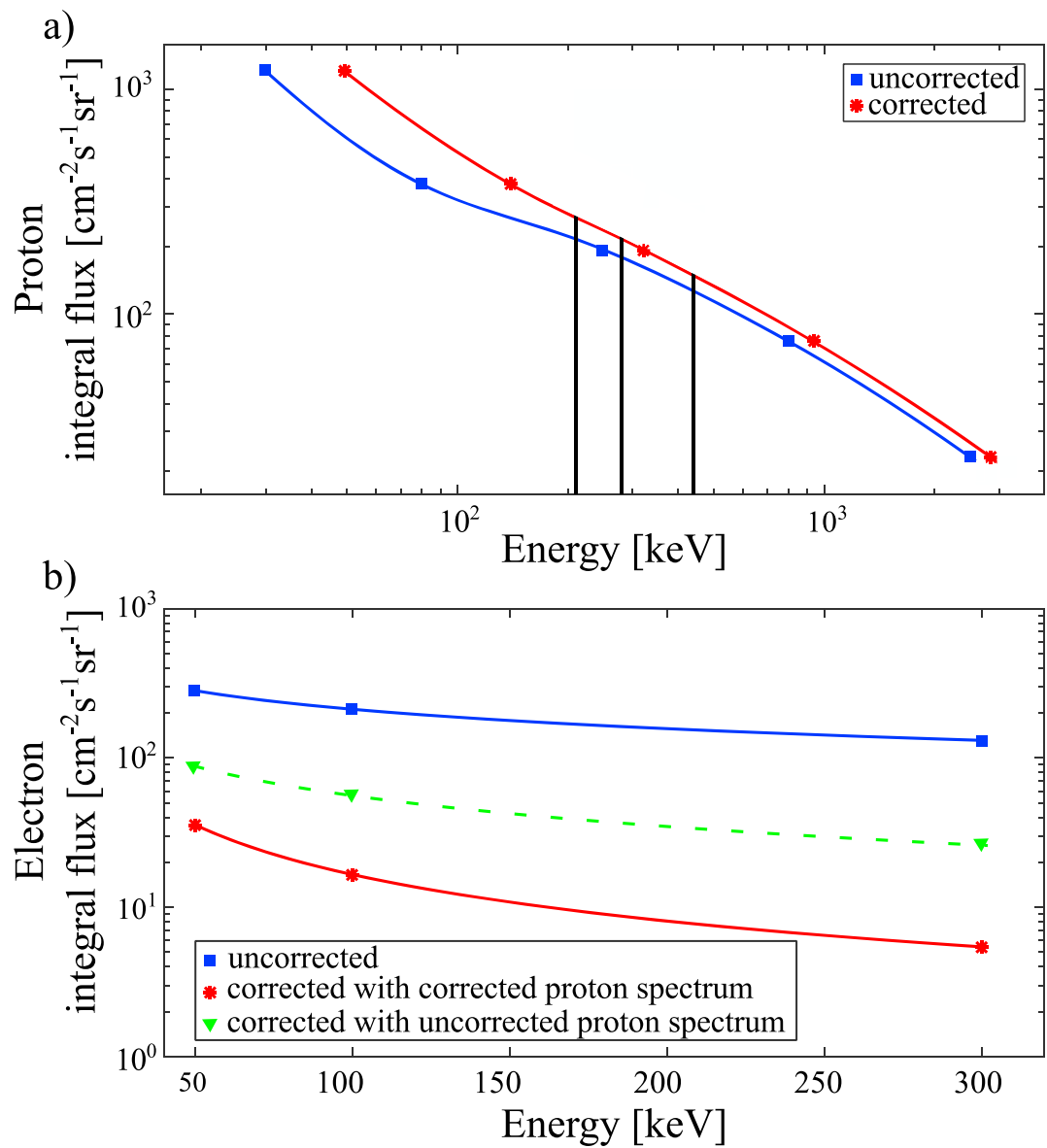

**Figure 2.** An illustration of proton and electron flux correction exemplified by NOAA/POES 15 in March 2013 at high latitudes in Northern Hemisphere. (a) The uncorrected integral spectrum (blue line) has the nominal energy thresholds: 30, 80, 250, 800, and 2500 keV. In the corrected spectrum (red line) the energy thresholds has changed according to the correction factors found by *Sandanger et al.* [2015]. The proton fluxes in the energy range from 210, 270, and 440 keV (marked as black vertical lines) to 2600 keV are subtracted from the original measured electron fluxes (blue line) in Figure 2b. The resulting fluxes represent the true electron fluxes in the respective energy bands (red line). The electron fluxes resulting from subtracting the uncorrected proton fluxes are also marked (green line).

The correction factors are applied to an integral flux spectrum to determine the pertinent energy ranges of the protons measured by the telescope used as described in *Sandanger et al.* [2015]. This is also illustrated in Figure 2a for a measurement by POES 15 in March 2012. The uncorrected integral flux spectrum (blue line, Figure 2a) has the nominal energy thresholds: 30, 80, 250, 800, and 2500 keV. In the corrected spectrum the energy thresholds have changed according to the correction factors, $\alpha_x$ found by *Sandanger et al.* [2015], $30\alpha_1$, $80\alpha_2$, $250\alpha_3$, $800\alpha_4$, and $2500\alpha_5$ keV (red line, Figure 2a). Subsequently, a monotonic Piecewise Cubic Hermite Interpolating Polynomial (PCHIP) is applied to the measured fluxes, and the proton fluxes in the energy ranges known to impact the respective electron channels are then retrieved and subtracted from the original measured electron fluxes (blue line, Figure 2b). The resulting fluxes represent the true electron fluxes in the respective energy bands (red line, Figure 2b). The importance of the correction of the proton flux spectrum is demonstrated in Figures 9 and 10 in *Sandanger et al.* [2015]. It is also demonstrated in Figure 2b. The uncorrected proton fluxes will cause a systematic overestimation of the electron fluxes (green line). The PCHIP routine is similar to the interpolating method presented by *Asikainen and Mursula* [2011], but we use different correction factors for the proton degradation. Other studies attempting to remove the proton contamination have used slightly different approaches in terms of creating a differential proton spectrum. *Lam et al.* [2010] estimated the proton energy spectrum by fitting a series of piecewise exponential functions across each measured proton energy channel range combined with the bow tie method [*Selesnick and Blake*, 2000]. *Peck et al.* [2015] performed various fits to the spectral distribution: exponential, power law, single Maxwellian, and double Maxwellian. *Lam et al.* [2010] and *Peck et al.* [2015], however, did not account for the proton detector degradation. They will subsequently overestimate the resulting electron fluxes in the cases when the proton contamination is present. The degree of overestimation will depend on the degradation of the proton detectors and the proton fluxes measured.

## 4. Capturing the Relativistic Electrons >1 MeV With the MEPED Proton Telescopes

There is also contamination from relativistic electrons in the proton fluxes. The collimator magnets sweep aside the electrons less than ~1 MeV. Higher-energy electrons, however, will hit the detector and possibly be detected as protons in P1, P2, P3, or P6, while P4 and P5 have a small responsivity to relativistic electrons. At electron energies >1 MeV the P6 is highly sensitive to electrons and parity with the proton response is reached at ~2 MeV [*Yando et al.*, 2011].

Previously, this feature has been applied in qualitative studies of relativistic electrons [*Miyoshi and Kataoka*, 2008; *Sandanger et al.*, 2007, 2009; *Rodger et al.*, 2010]. The geometric factors provided by *Yando et al.* [2011] enable a quantitative assessment of the fluxes. *Peck et al.* [2015] utilize the estimated geometric factors in a sophisticated proton correction routine and apply the geometric factors of differential energies consistent with Appendix B in *Yando et al.* [2011]. The differential energy spectra are estimated by fitting the measurements of P1–P5 to energy exponential, power law, single and double relativistic Maxwellian distributions, or a weighted combination of these. The respective spectral shapes are then extrapolated to higher energies. The discrepancies between the extrapolated fits and the fluxes measured in P6 are then assumed to be primarily relativistic electrons >1 MeV. The resulting MEPED electron fluxes are further applied in an in-flight comparison with electron fluxes measured by the IDP (Instrument for Detecting Particles) on DEMETER (Detection of Electromagnetic Emissions Transmitted from Earthquake Regions) spacecraft [*Sauvaud et al.*, 2006]. The comparison shows that electron fluxes measured by MEPED are greater than electron fluxes measured by IDP, which *Peck et al.* [2015] attribute partially to differences in pitch angle range for the measurement on the two spacecraft. As previously mentioned, using uncalibrated proton spectra might also contribute to an overestimate of the electron fluxes measured by MEPED.

The routine by *Peck et al.* [2015] offers an estimate of >1 MeV fluxes continuously, only disrupted by solar proton events and the South Atlantic anomaly (SAA). We apply a somewhat simpler approach to avoid an overestimate of the relativistic electron fluxes. We use the fluxes reported in the P6 channel only when there are little or no protons detected in the P5 channel. We require the count rate in P5 to be less than 10% of the count rate in P6. This automatically excludes regions impacted by SPEs, geomagnetic storms accelerating protons to high energies, and the SAA. Further, we demand that the P6 channel fluxes are lower than the proton corrected E3 channel fluxes. The count rate in P6 is then assumed to be solely due to >1 MeV electrons.

Figure 3 shows the differential proton fluxes for P4 (800–2500 keV) and P5 (2500–6900 keV), the mixed proton and electron flux response in P6 (>6900 keV protons/> 1000 keV electrons), and the estimated pure relativistic electron response in P6 in a year of solar maximum (2003) and a year of solar minimum (2008). The P5 channel has in general low or no counts. This implies that we will have a reasonable continuous coverage of the >1000 keV radiation belt electrons throughout the solar cycle as shown in Figure 3 (fourth and eight panels). It is therefore possible to investigate their general impact upon the atmosphere. The extensive time coverage of the POES and MetOp satellites will enable a quantitative long-term study of the behavior of these high-energy electrons. For the year 2003 there are, however, multiple time periods of enhanced fluxes in the P5 channel. It is clear from Figure 3 (fourth panel) that these periods are efficiently eliminated by the applied criteria. Although radiation belt electrons generally will be accelerated in these geomagnetic storms, the intensity level cannot be retrieved from MEPED during periods of high fluxes of protons hitting the detectors.

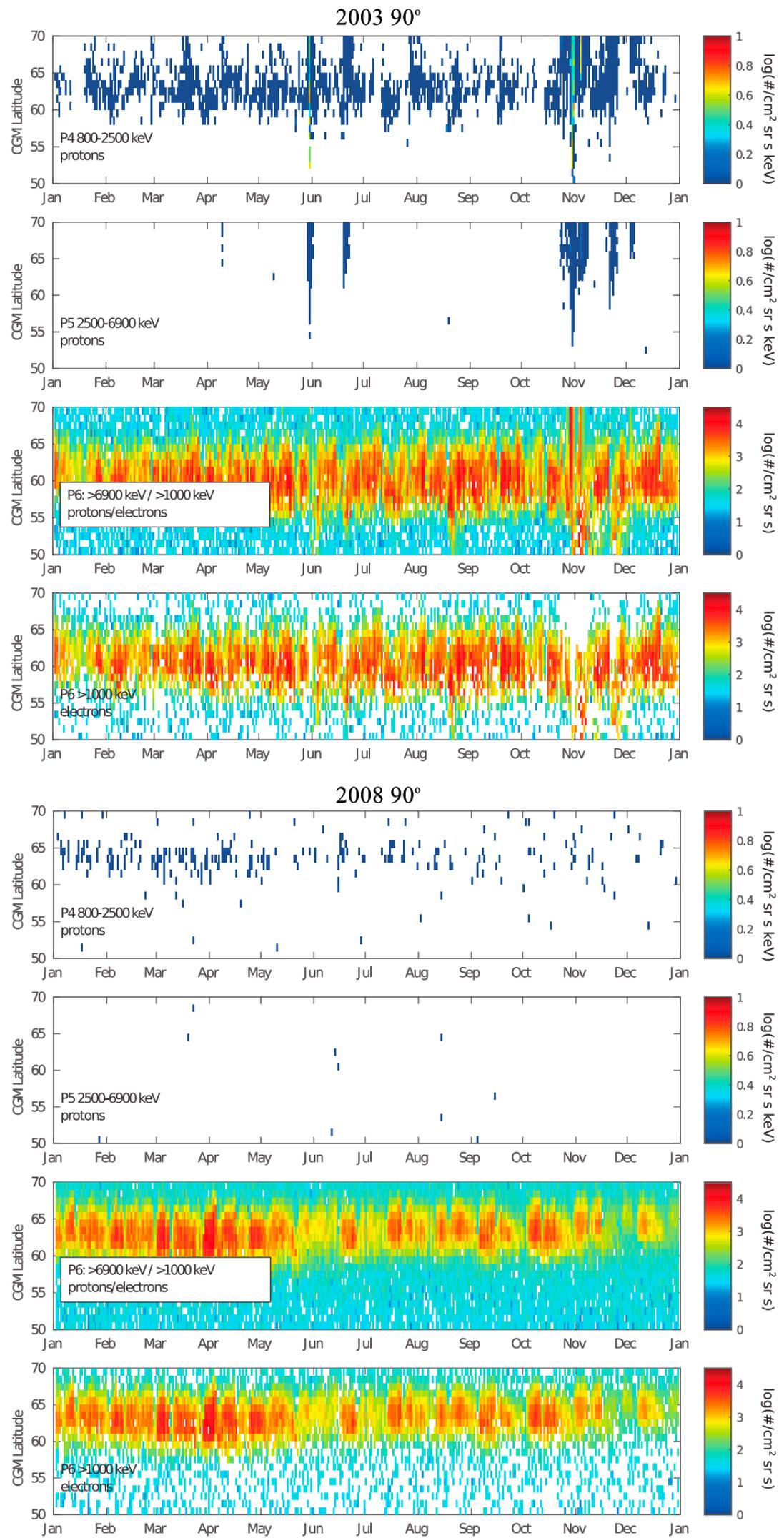

**Figure 3**

On the other hand, due to the nature of the electron acceleration processes, it is expected that they will be delayed in respect to the onset of the geomagnetic storms. It will therefore be possible to study their potential impact upon the atmosphere in the aftermath of even strong SPEs where high-energy electrons are clearly evident at latitudes down to about 50° (Corrected Geomagnetic) latitudes.

## 5. Determining the Level of Diffusion and the Loss Cone Pitch Angle Distribution

With the corrected and extended electron spectra, we can now determine the level of electron pitch angle anisotropy and diffusion using measurements from both the 0° and 90° telescopes in a combination with theoretically determined PADs. Taking into account the detector response for different PADs, the 0° and 90° fluxes are fitted onto the solution of the Fokker-Planck equation for wave-particle interactions [*Kennel and Petschek*, 1966]. Finally, we estimate the equivalent isotropic flux level over the bounce loss cone in order to give a more precise estimate of the energy deposition in the upper atmosphere.

### 5.1. Accounting for the Detector Geometry

*Yando et al.* [2011] provide geometric factors for the entire detector area given the simplifying assumption that the incident flux is isotropic. In most cases, except for very strong diffusion, the electron fluxes are anisotropic. This implies that the directional flux varies over the acceptance solid angle of the detector, see Figures 1a–1c for one case at high latitudes. The effective detector area also varies with the viewing angle through the collimator, see Figure 1d. The resulting detector count rate is derived by integrating the product over the solid angle of the detectors field of view. This is analogous to Figure 1b shown in *Li et al.* [2013] illustrating the geometry for calculating counting rates of the detector.

The offset between the direction of the telescopes and the magnetic field line where the PAD is centered varies over the orbit of the satellite. Hence, the detector sensitivity needs to be taken into account for different PADs and different pointing directions for the telescopes.

### 5.2. Quasi-Linear Theory of Wave-Particle Interaction

Pitch angle scattering by wave-particle interaction has received considerable attention from the very early stages of studying the dynamics of the radiation belt from both theoretical and experimental points of view. This effort resulted in the solution of the Fokker-Planck equation for particle diffusion and yields the following equilibrium equatorial PAD [*Kennel and Petschek*, 1966; *Theodoridis and Paolini*, 1967]:

$$j(E, D, \alpha_{eq}, B_{eq}, L) = N \cdot S(E) D^{-1} h(\alpha_{eq}) \tag{1a}$$

$$j(E, D, \alpha_{eq}, B_{eq}, L) = N \cdot S(E) D^{-1} \left[ h(\alpha_0) + \ln\left(\frac{\sin\alpha_{eq}}{\sin\alpha_0}\right) \right] \tag{1b}$$

inside $(\alpha_{eq} \leq \alpha_0)$ and outside $\left(\alpha_0 \leq \alpha_{eq} \leq \frac{\pi}{2}\right)$ the loss cone, respectively, where

$$h(\alpha_{eq}) \equiv \frac{\sqrt{DT}}{\alpha_0} \left[ \frac{I_0\left(\frac{\alpha_{eq}}{\sqrt{DT}}\right)}{I_1\left(\frac{\alpha_0}{\sqrt{DT}}\right)} \right]$$

$j(E, D, \alpha_{eq}, B_{eq}, L)$ is the equatorial distribution for electrons of energy, $E$, diffusion coefficient $D$, and pitch angle, $\alpha_{eq}$, at the equatorial point, $B_{eq}$. $N$ is a normalizing constant. $S(E)$ is the particle injection source, assumed to be nonzero only at $\alpha_{eq} = \frac{\pi}{2}$. The $\alpha_0$ is the equatorial loss cone angle. $T$ is the escape time of a particle that is inside the bounce loss cone. $I_0$ and $I_1$ are the modified Bessel functions of first kind. The diffusion coefficient is defined as

$$D \equiv |\cos\alpha_{eq}| \frac{\Delta\alpha_{eq}^2}{\delta t}$$

The above equations do not depend on the specific mechanism responsible for the pitch angle diffusion [*Theodoridis and Paolini*, 1967]. Their validity is, however, limited to steady state diffusion. Nevertheless, recent results by, e.g., *Li et al.* [2013, 2014a, 2014b] have validated the application of the Fokker-Planck equations by analyzing conjunction events between the POES/MEPED electron observations and the chorus wave amplitudes measured by the Van Allen Probes during multiple geomagnetic storms with strong electron flux variation. *Kennel and Petschek* [1966] also pointed out that the shape of the PAD outside the loss cone is essentially independent of the magnitude of the diffusion coefficients as long as it is nonzero. At high latitudes the loss cone constitutes just a marginal part of the fluxes, so it is therefore not likely to affect the general PADs at short timescales.

**Figure 3.** Pass-by-pass MEPED 90° detector P4, P5, and P6 fluxes binned in a uniform grid of CGM latitude and date of 1° and 1 day for 18–24 MLT for the years 2003 and 2008. Panels 1,2, 5 and 6 show P4 (800–2500 keV) and P5 (2500–6900 keV) differential fluxes multiplied with the respective energy band width. Panels 3 and 7 give the uncorrected P6 integral fluxes (>6900 keV protons/>1000 keV electron fluxes). Panels 4 and 8 give the estimated >1000 keV electron fluxes.

Taking advantage of the relatively large loss cone at ~850 km compared to the equator, we transform the entire PAD from the equator to the satellite altitude and thereafter to the top of the atmosphere (~120 km) presuming that the pitch angles are redefined in terms of the magnetic field at any location by the first adiabatic invariant:

$$\frac{|v|\sin^2\alpha}{B} = \text{constant} \tag{2}$$

and assuming that $|v|$ is constant along the path. This assumption will not hold for low-energy electrons where the acceleration by parallel electric field needs to be taken into account, but it is approximately valid for the MEPED energy ranges > 50 keV. We then get

$$\sin^2\alpha = \sqrt{\frac{B}{B_{eq}}}$$

where $B_{eq}$ is the magnetic flux density at the equator.

All the transformations of the PADs for different levels of diffusion strength, $\sqrt{DT} = \frac{1}{X}$, have been done in advance, creating an extensive library of look-up tables where the denominator changes in steps of 1 from 1 to 400 and thereafter in steps of 10 from 400 to 4000. They apply to all possible loss cone sizes and their respective PADs. All sets of PAD are made at equator, at satellite altitude, and at the top of the atmosphere (~120 km).

We estimate the loss cone size at the equator for all possible latitude and longitude positions of the satellite, by tracking along the magnetic field lines using the IGRF model. The result is an additional look-up table. When applying the routine to electron flux measurement at a specific location of the satellite (at one specific data point), we use this look-up table to find the specific loss cone size, which then is used to find the correct set of PADs in the library. When comparing the theoretical PADs with the measured particle fluxes the procedure is as follows:

1. Determine the pitch angles of center look directions of the 0° and 90° telescopes.
2. Calculate the ratio, $R_O$, between of the fluxes detected by the 0° and 90° detector.
3. Calculate the ratio, $R_T$, between the fluxes of the two detectors for theoretical PADs using the procedure described in section 5.1. This is done for a dense set of PADs corresponding to different values of the diffusion strength, $DT$, using the relevant look-up table.
4. Determine which PAD gives the same ratio $R_T$ as the observed ratio, $R_O$. This gives us the right diffusion level and the specific PAD of the observed particle fluxes is determined.

Finally, from the PAD at 120 km, we calculate the number of particles that crosses a unit horizontal area per second. We then find the equivalent isotropic flux which gives the same number of particles per second through the unit horizontal area, which we refer to as the equivalent isotropic flux level over the bounce loss cone. Each energy interval is treated separately as the level of diffusion will depend on the particle energy. A simplified summary of the steps of the procedure are illustrated in Figure 4.

## 6. The Electron Precipitating Fluxes During the HSSWS of February 2008

In order to check if our procedure leads to a realistic estimate of the precipitating electron fluxes, we investigate the OH response to a weak geomagnetic active period driven by HSSWS in the beginning of February 2008. The solar wind starts to increase around 12 UT of 31 January reaching 600 km/s around 15 UT on 1 February as shown in Figure 5. The elevated solar wind speed is sustained for almost 3 days accompanied with moderate *Ap* values of 17–19. The *AE* index does also show moderate deflection reaching a maximum level of approximately 900 nT. Based on the classification of geomagnetic storms by *Loewe and Prölss* [1997], this event qualifies as a weak geomagnetic storm as the *Dst* reaches two minimum values about −30 and −50 nT. The double structure is most likely due to the plateau seen in the solar wind speed in the same period.

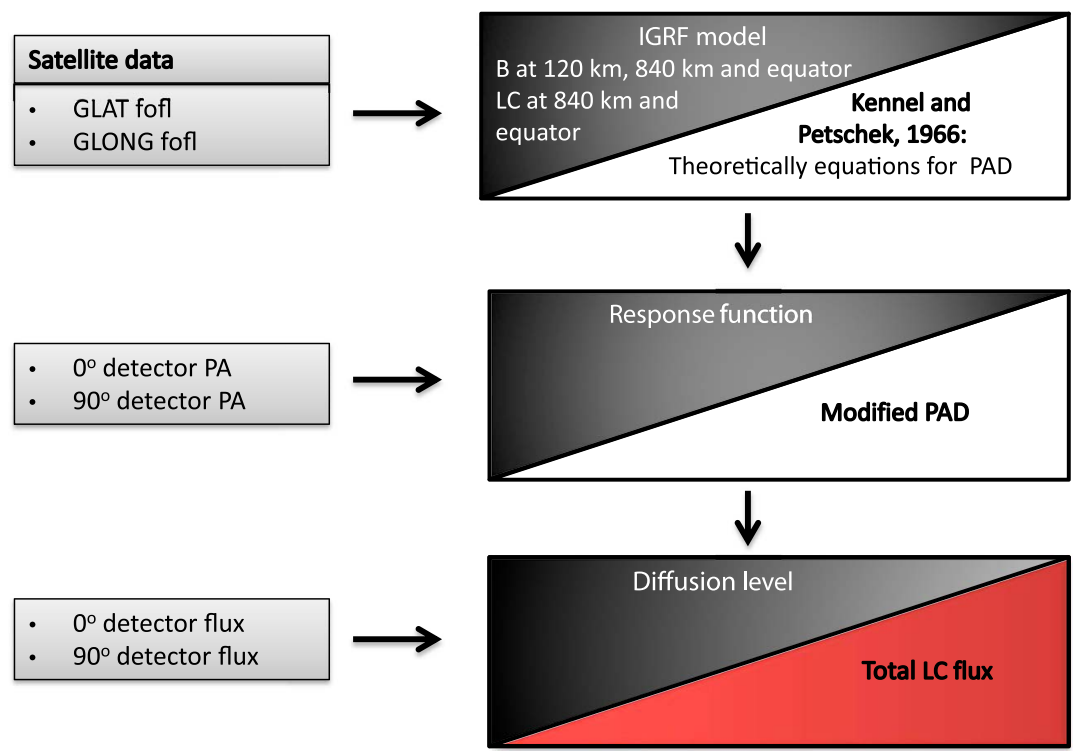

**Figure 4.** An overview of the procedure demonstrating the use of satellite data together with the IGRF model, theory of wave-particle interactions in the magnetosphere, and the angular response function of the detectors. The result is an estimate of the diffusion level and subsequently the total loss cone fluxes.

Figure 6 shows that the $> 50$ keV trapped electron fluxes for 62°–64° CGM latitude measured by the 90° detector increases as the $Dst$ decreases. The fluxes are still elevated at the end of the storm indicating that the storm causes enhancement of the radiation belt fluxes. A similar, but delayed response, is seen in the $> 100$ keV, $> 300$ keV, and $> 1000$ keV trapped fluxes. This is in line with current theory suggesting that the acceleration processes associated with potential relativistic electrons in the radiation belt requires both a seed population and time [e.g., *Summers et al.*, 1998; *Horne et al.*, 2009]. The electron fluxes measured by the 0° detector, on the other hand, show an elevated, but more variable behavior more in line with the *AE* index. For the $>50$ keV fluxes the maxima reach almost the level of the trapped fluxes indicating strong pitch angle diffusion, but in general, the fluxes appear strongly anisotropic. At higher energies the ratios between the 90° and 0° fluxes indicate weaker pitch angle diffusion.

In the cases of strong anisotropy it is necessary to require a lower limit for the 0° electron fluxes. If the 0° detector report fluxes close to the noise level, we cannot determine its real value. This will overestimate the pitch angle diffusion strength and hence lead to an overestimation of the loss cone flux. Thus, we only infer the loss cone flux where the $> 50$ keV and $> 100$ keV fluxes measured by both 0° and 90° are sufficiently larger than the background level ($\sim 300 \, cm^{-2} s^{-1} sr^{-1}$) and assume zero fluxes otherwise. This is more prudent compared to *Rodger et al.* [2013] which uses fluxes of $100 \, cm^{-2} s^{-1} sr^{-1}$ as the lower limit, but weaker than the limit $500 \, cm^{-2} s^{-1} sr^{-1}$ applied by *Li et al.* [2013]. In practice, this implies that our routine will be unfit to determine the medium electron energy drizzle into the atmosphere at the current spatial and temporal resolution. Assuming zero fluxes in these cases will therefore potentially underestimate of the precipitating fluxes prior to and after an event.

In 2008 NOAA/POES 18 share the same local time coverage as the Aura satellite which measures the OH density. This means that we can determine the precipitating fluxes close in both time and space to the Sun-synchronous polar orbiting Aura satellite. We utilize the OH observations measured by the Aura/MLS instrument. The data are version 3.3, level 2 nighttime (solar zenith angle $>100°$) and are screened according to the selection criteria by *Livesey et al.* [2013]. The OH concentration is given as parts per billion which we convert to number density by also utilizing the MLS temperature measurements.

The potential impact of particles on the OH production will be oriented in geomagnetic coordinates. Evaluating OH in a geomagnetic coordinate system will, however, impose a daily variation in OH. This is due to the offset between the two coordinate systems as several processes such as tides, planetary waves, photolysis, and other species interacting with OH, are better lined up in geographic coordinates. To avoid this diurnal variation in CGM coordinates, we use a daily running mean at the specific CGM latitude. The result is shown in Figure 7a for the altitude range 64–78 km. The altitude range is chosen based on the energy deposition during this event (see Figure 7b). Above 80 km there is not sufficient pressure to form water cluster ions needed in the EEP-HO$_x$ production [*Solomon et al.*, 1981; *Sinnhuber et al.*, 2012].

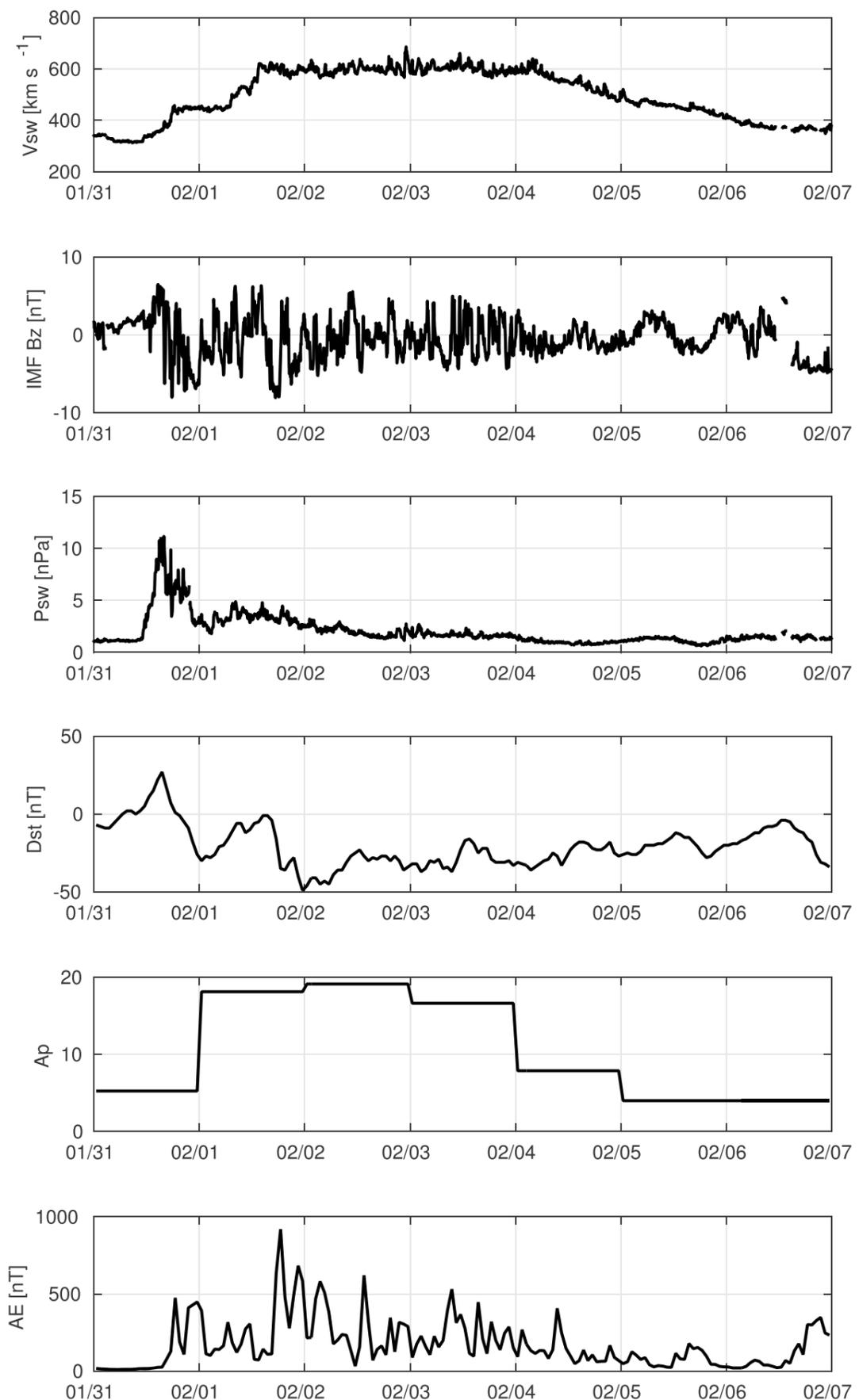

**Figure 5.** Solar wind conditions and geomagnetic indices on 31 January to 6 February 2008.

The first geomagnetic quiet days illustrate a strong altitude gradient in the OH density. One of its key features of nighttime OH is the presence of an enhancement of OH in a narrow layer around 82 km [*Pickett et al.*, 2006]. It is linked to collisional quenching of vibrational excited OH that is caused by the reaction of ozone with atomic hydrogen [*Damiani et al.*, 2010]. In Figure 7a the OH density gradually decreases toward a local minimum around 75 km. Here the main source of OH is photolysis of water vapor by Lyman alpha. During winter at high latitudes (~60°N/S) in the altitude range of 66–77 km, near the polar night terminator, however, the nearly grazing incidence of solar radiation makes the atmosphere optically thick leading to attenuation of this radiation. This will result in low background HOx production and a minimum in the background OH in this altitude range [*Marsh et al.*, 2001; *Sonnemann et al.*, 2006]. Similar features of a local OH minimum are also found by, e.g., *Damiani et al.* [2010].

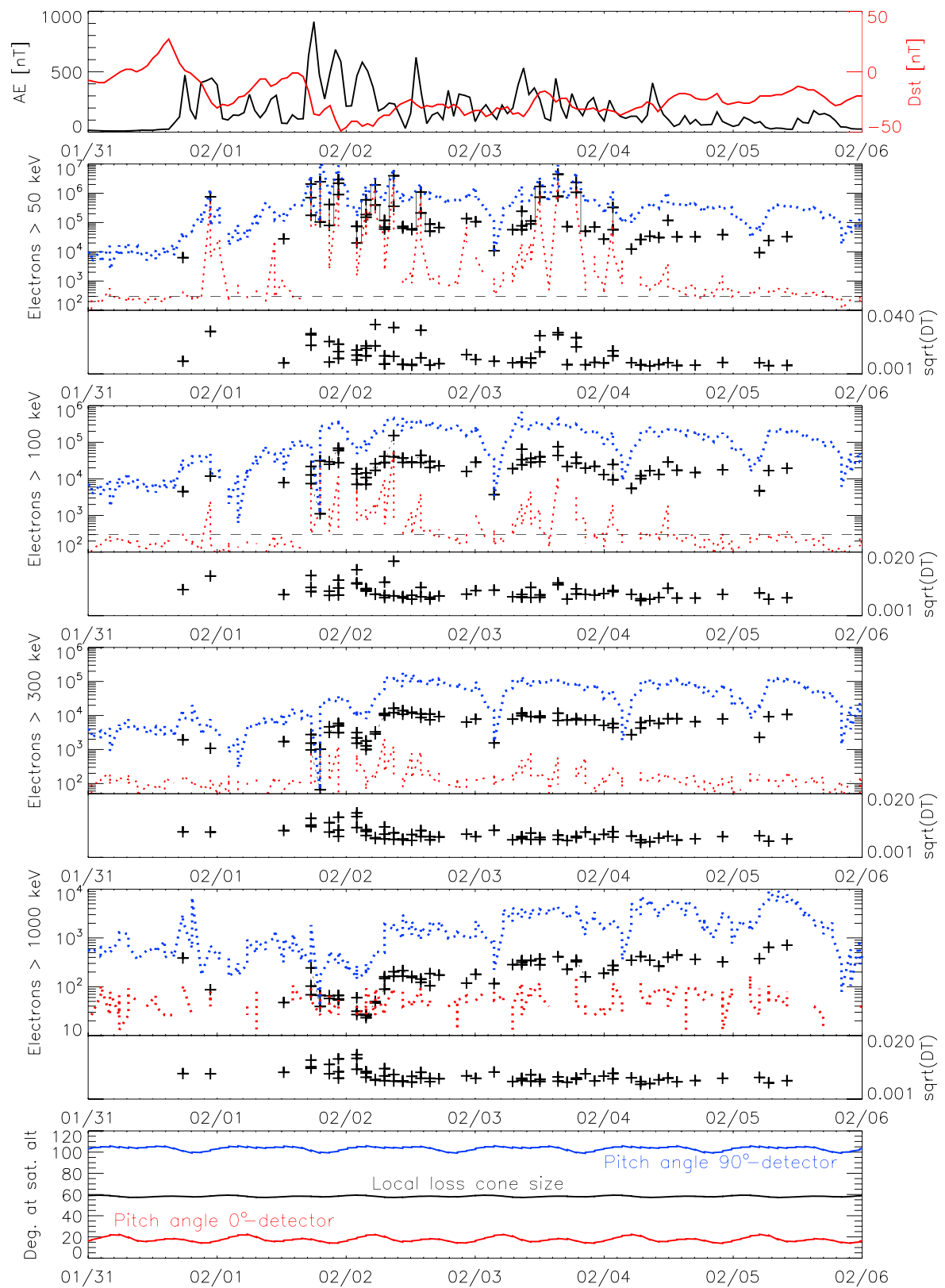

**Figure 6.** (first panel) *AE* (black) and *Dst* (red) from 31 January to 6 February 2008. (second to fifth panels) Electron fluxes >50, >100, >300, and >1000 keV measured by the 0° (red dotted line) and 90° (blue dotted line) detector on NOAA/POES 18 from 31 January to 6 February 2008 at 62–64 CGM latitude for nighttime conditions. The estimated equivalent isotropic fluxes in the loss cone and the associated diffusion strength, $\sqrt{DT}$, are marked as black crosses. The black dashed lines in the two uppermost electron flux panels illustrate the criteria applied for estimating the loss cone fluxes. (sixth panel) Loss cone size variability (black) and pitch angle pointing direction of the 0° (red) and 90° (blue) telescopes.

Using the estimate of the equivalent isotropic flux level over the bounce loss cone, we can utilize the cosine-dependent IDH (Isotropic over the Downward Hemisphere) model of *Rees* [1989] to calculate the subsequent energy deposition as a function of altitude. This is based on a standard reference atmosphere (Committee on

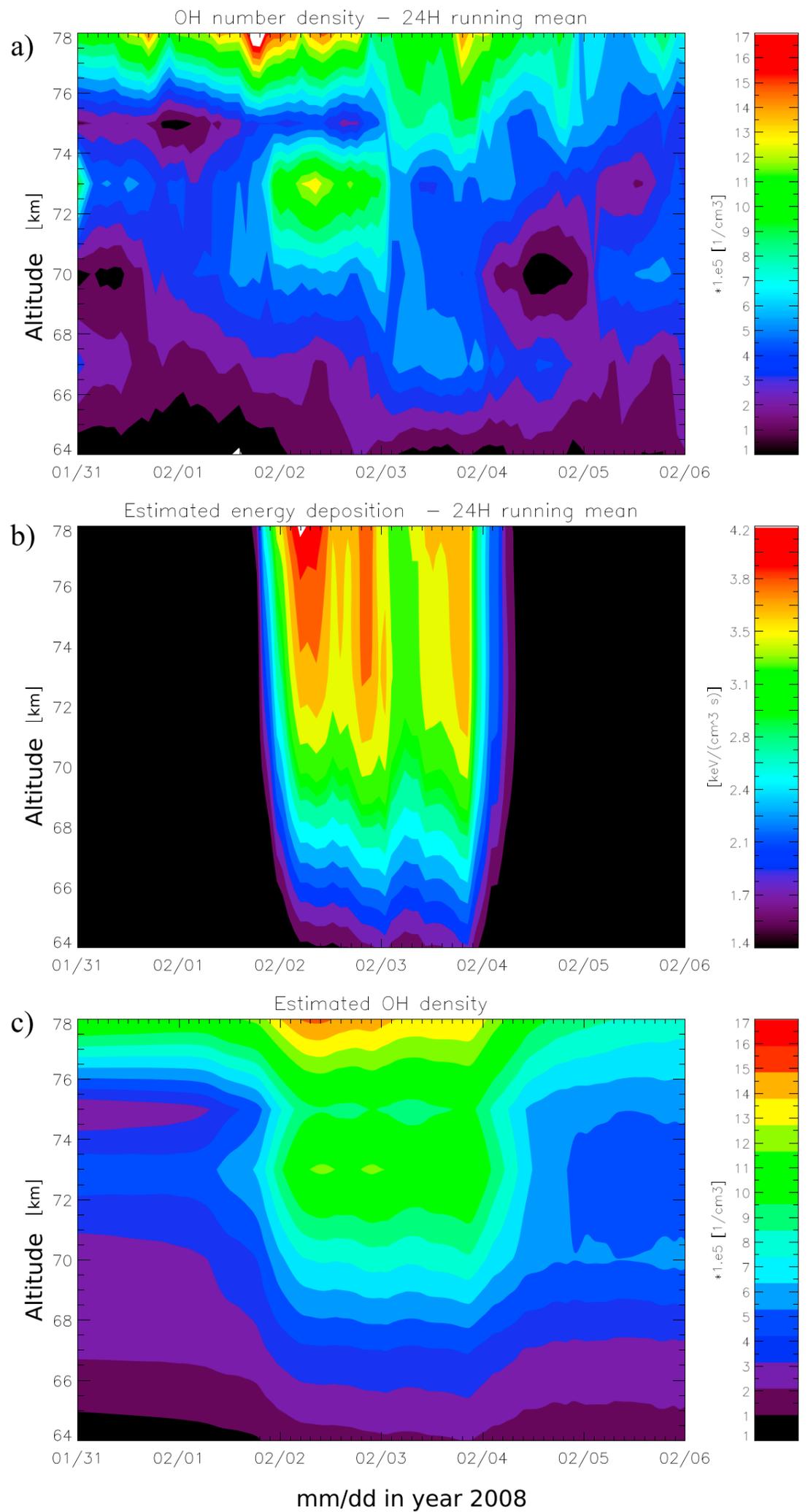

**Figure 7.** (a) The 24 h running mean of the nighttime OH measured by MLS AURA at 62°–64° CGM latitude at the Northern Hemisphere from 31 January to February 2008. For the same period, (b) the estimated energy deposition based on the loss cone flux method for MEPED electron measurement from NOAA/POES 18. (c) The energy deposition estimate is then used in a simple model to predict the expected OH density.

Space Research International Reference Atmosphere 1986). The result is shown in Figure 7b for a 24 h running mean. In Figure 7a we see that the OH density increase significantly beyond the background at all altitudes from 67 to 78 km around 12 UT on 1 February, maximizing around 10 UT on 2 February. At this point of time the OH density is elevated in the entire altitude interval from 64 km to 78 km. Both the time and altitude change are well aligned with the increased fluxes and the estimated energy deposition as function of altitude. The local extremes seen in the background OH are still visible, but the local minimum at 75 km altitude is less pronounced. The energy deposition also clearly displays a secondary maximum during 3 February. This is weaker but still distinctly evident also in OH. The elevated OH density between 64 and 68 km appears to have a stronger response to the energy deposition compared to the response above 70 km. Qualitatively, it is clear that the electron precipitation even during a weak geomagnetic storm can cause a measurable increase in OH also in the lower mesosphere, here down to about 64 km.

To quantitatively evaluate the impact of the electron fluxes, we apply a simple model making a rough approximation of the OH production and loss. We assume that the electron fluxes and the subsequent energy deposition are linearly changing between each passage (~100 min) with step size of 5 min. We adopt a height-dependent OH production rate based on the model results from the University of Bremen Ion Chemistry model (UBIC) shown in Figure 6 in *Sinnhuber et al.* [2012]. At 70 km and below we assume a production rate of 0.9 OH molecules per ion pair, while at 73, 75, and 78 km the production efficiency is reduced to 0.75, 0.65, and 0.4 OH per ion pair, respectively. We assume a height-dependent lifetime monotonously varying from 0.5 to 3 h in the altitude interval 64–78 km. The OH production from EEP and the subsequent loss is treated separately from the background OH. We assume a background OH concentration based on an average of 1 day prior and 1 day after the elevated electron fluxes requiring a smooth linear transition throughout the event. The result is shown in Figure 7c. It is evident that the estimated OH density is in the same order of size as the observed OH density. In particular, the first part of the event appears to be well in line with the OH estimate on all height levels. The later part of the event appears to overestimate the level of OH in the altitude region between 70 and 75 km. Considering the OH increase found below 70 km appear to be related to the energy deposition, we speculate if the lack of agreement between 70 and 75 km can be related to short-time changes in background level of OH such as gravity wave activity. The EEP-OH production efficiency may also depend on the level of atomic oxygen and water vapor [*Solomon et al.*, 1981], which will vary with the dynamics of the background atmosphere. All in all, considering a highly varying atmospheric background and the simplicity of our OH production model, we find it to be probable that our loss cone fluxes give a realistic estimate of the true precipitating electron fluxes. In contrast, applying just the 0° or the 90° detector fluxes will underestimate or overestimate the OH production by an order of magnitude (not shown).

## 7. Conclusion and Future Work

In order to achieve a realistic estimate of the precipitating electron fluxes, we apply the measurement from both the 0° and 90° telescopes on the MEPED detector. We correct for proton contamination in the electron counts, where the proton spectra have already been corrected due to detector degradation. We also account for the relativistic (>1000 keV) electrons contaminating the proton detector. This gives us full range coverage of electron energies that will be deposited in the middle atmosphere from 60 to 80 km. By combining the measurements from both telescopes with electron pitch angle distributions from theory of wave-particle interactions in the magnetosphere, a complete bounce loss cone flux is constructed for each of the electron energy channels > 50 keV, > 100 keV, > 300 keV, and > 1000 keV. The result should be considered with some caution as there will be times when the PADs deviates considerably from steady state diffusion.

Comparing the estimated fluxes and subsequent energy deposition to the OH density as measured by the MLS on board the Aura satellite during a weak geomagnetic storm, we substantiate that our method gives a realistic estimate of precipitating energetic electron fluxes. We find that the estimated fluxes both qualitatively and quantitatively are in line with the observed increase of $HO_x$ during a weak geomagnetic storm. It should, however, be noted that this is not an infallible validation considering the assumptions of time variations, OH background, etc. Further validation of the method could be achieved with, for example, comparison with indirect measurements of the particle precipitation such as cosmic radio noise absorption measured by riometers. More work should also be done to achieve also a good estimate of the potential electron drizzle when the 0° detector measures close to the noise floor. In addition, assessment of the limitation and subsequent errors associated with the use of the IGRF model should be made for varying levels of geomagnetic activity.

Quantifying of the level of precipitating energetic electrons is essential in order to determine the level of EEP production of OH and NO in the middle atmospheric chemistry and hence its potential impact upon ozone, temperature, and dynamics. The presented method can be applied to all NOAA/POES and MetOp satellites and with their wide local time coverage the measurements could be combined to achieve a global estimate of the electron energy input in the atmosphere during periods of enhanced geomagnetic activity.

**Acknowledgments**
This study was supported by the Research Council of Norway under contracts 184701 and 223252/F50. The authors thank the NOAA's National Geophysical Data Center (NGDS) for providing NOAA data (http://satdat. ngdc.noaa.gov/) and WDC Geomagnetism, Kyoto, Japan, for AE and Dst indices (http://wdc.kugi.kyoto-u.ac.jp/wdc/Sec3.html); SPDF Goddard Space Flight Center for solar wind parameters (http://omniweb.gsfc.nasa.gov/); and NASA Goddard Earth Science Data and Information Services Center (GES DISC) for providing Aura/MLS data (http://mls.jpl.nasa.gov/).